\documentclass[sigconf]{acmart}
\AtBeginDocument{%
  }
\usepackage{amsmath,amsfonts}
\usepackage{algorithmic}
\usepackage{array}
\usepackage{balance}
\usepackage[caption=false,font=normalsize,labelfont=sf,textfont=sf]{subfig}
\usepackage{textcomp}
\usepackage{stfloats}
\usepackage{url}
\usepackage{cite}
\usepackage{verbatim}
\usepackage{graphicx}
\usepackage{nicematrix}
\usepackage{balance}
\usepackage{multirow}
\usepackage{enumerate}
\usepackage[utf8]{inputenc}
\usepackage{graphicx}
\usepackage{float}
\usepackage{hyperref}
\usepackage{adjustbox}
\usepackage{enumitem}
\usepackage{booktabs}
\usepackage{caption}
\usepackage{xcolor}
\usepackage[table]{xcolor}
\usepackage{algorithm}
\usepackage{algorithmic}
\usepackage{soul} 
\usepackage{tcolorbox}
\tcbuselibrary{skins, breakable}
\usepackage{mathtools}
\usepackage[normalem]{ulem}

\begin{document}
\renewcommand\footnotetextcopyrightpermission[1]{}
\title{MLLMRec-R1: Incentivizing Reasoning Capability in Large Language Models for Multimodal Sequential Recommendation}

\author{Yu Wang}
\orcid{0009-0008-6272-8714}
\affiliation{%
  \institution{Hefei University of Technology}
  \city{Hefei}
  \country{China}}
\email{wangyu20001162@gmail.com}

\author{Yonghui Yang}
\affiliation{%
  \institution{National University of Singapore}
  \city{Singapore}
  \country{Singapore}}
\email{yyh.hfut@gmail.com}

\author{Le Wu}
\affiliation{%
  \institution{Hefei University of Technology}
  \city{Hefei}
  \country{China}}
\email{lewu.ustc@gmail.com}

\author{Jiancan Wu}
\affiliation{%
  \institution{University of Science and Technology of China}
  \city{Hefei}
  \country{China}}
\email{wujcan@gmail.com}

\author{Hefei Xu}
\affiliation{%
  \institution{Hefei University of Technology}
  \city{Hefei}
  \country{China}}
\email{hefeixu604@gmail.com}

\author{Hui Lin}
\affiliation{%
  \institution{China Academy of Electronic and Information Technology}
  \city{Beijing}
  \country{China}}
\email{linhui@whu.edu.cn}

\renewcommand{\shortauthors}{Trovato et al.}

\begin{abstract}
Group relative policy optimization (GRPO) has become a standard post-training paradigm for improving reasoning and preference alignment in large language models (LLMs), and has recently shown strong effectiveness in LLM-based recommender systems. However, extending GRPO-based reasoning pipelines to multimodal sequential recommendation (MSR) with multimodal large language models (MLLMs) faces fundamental obstacles. First, MSR requires jointly encoding visual content for both historical interactions and multiple candidate items, causing visual tokens to dominate the input and making the cost of group-based rollout scale with history length and candidate set size, which renders GRPO-based training prohibitively expensive. Second, existing Chain-of-Thought (CoT) supervision suffers from reward inflation in recommendation scenarios, where higher training rewards do not reliably translate into improved ranking performance and may induce shortcut learning. To address these challenges, we propose MLLMRec-R1, an efficient and stable GRPO-based reasoning framework for multimodal sequential recommendation. MLLMRec-R1 textualizes visual signals offline to eliminate expensive visual tokens while preserving multimodal semantics, and constructs high-quality multimodal CoT supervision through refinement and confidence-aware assessment. Furthermore, a mixed-grained data augmentation strategy selectively injects reliable CoT samples while retaining standard training data, mitigating reward inflation and improving generalization stability. Extensive experiments on three benchmark datasets demonstrate that MLLMRec-R1 consistently outperforms state-of-the-art methods, establishing a practical and effective GRPO-based reasoning pipeline for multimodal sequential recommendation. The code is available at this link\footnote{https://github.com/wangyu0627/MLLMRec-R1}.
\end{abstract}




\maketitle

\section{Introduction}

\begin{figure}[t]
    \centering
    \includegraphics[width=\linewidth]{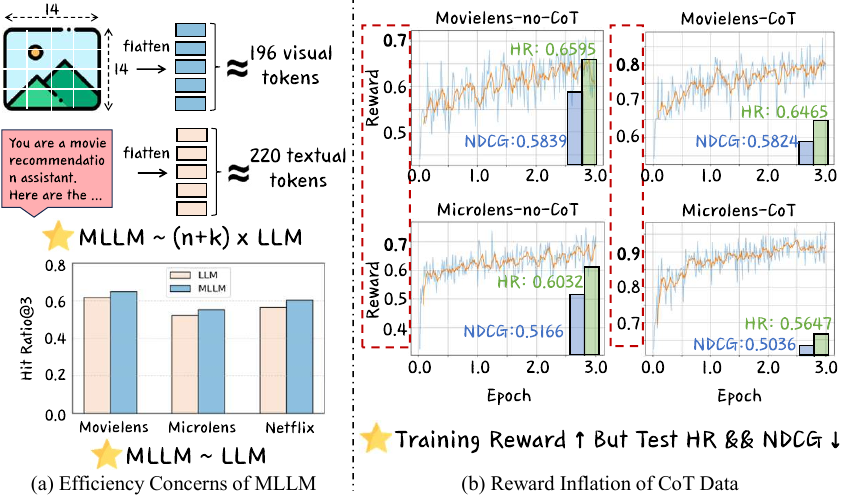}
    \caption{(a) Visual tokens make compute grow with history length and candidate set size, yet often yield limited gains over LLMs, revealing an efficiency bottleneck. (b) Some CoT data may boost training reward scores but hurt test performance, revealing shortcut learning and poor generalization.}
    \label{fig:intro}
    \vspace{-0.3cm}
\end{figure}

With the rapid growth of short-video and e-commerce platforms, multimodal sequential recommendation (MSR)~\citep{2024msr, 2025multimodal, 2023mmmlp} leverages multimodal signals (e.g., images~\citep{2023MoRec}, audio~\citep{2023mmssl}, and videos~\citep{2019mmgcn}) to infer users’ future interests. Benefiting from the strong multimodal understanding and generation capabilities of Multimodal Large Language Models (MLLMs)~\citep{Qwen3-VL, 2024mdpo, 2024LLaVA-RLHF}, recent studies~\citep{2025hanorec, 2025MLLM-MSR, 2025MSRBench} increasingly adopt them to improve MSR performance. However, recommendation-oriented inference differs from general question answering~\citep{2025OPA-DPO}: MSR must handle long interaction histories and multiple candidates in structured inputs $\langle \textit{id}, \textit{title}, \textit{image} \rangle$. As shown in Figure~\ref{fig:intro} (a), visual encoding turns a single image into 196 visual tokens, often comparable in scale to the text prompt. The token count grows roughly linearly with the history length and candidate set size, leading to high memory use and inference latency. In many cases, the gains over text-only models do not justify this cost. \textbf{Can we represent visual semantics more efficiently to enable scalable training and inference for MLLM-based MSR?}

A natural approach is to compress visual signals offline into text, allowing the model to gain multimodal benefits at near text-only token cost during training and inference~\citep{2025vragent, 2025MLLM-MSR}. Furthermore, turning multimodal information into stable recommendation decisions remains challenging. Prior work~\citep{2023tallrec, 2025SPRec, 2025reczero, 2025collm} applies supervised fine-tuning (SFT)~\citep{2025-sft-1, 2025-sft-2, 2022lora} and preference learning such as direct preference optimization (DPO)~\citep{2023dpo, 2025SPRec, 2025hanorec} to align outputs, but these methods mainly fit final responses and often fail to organize multimodal evidence from history for fine-grained preference reasoning~\citep{2025llmrec-survey}. Group Relative Policy Optimization (GRPO)~\citep{2025-deepseek-r1, 2025reczero, 2025-DrGRPO} instead samples multiple responses under the same context and updates them relatively, encouraging exploration of reasoning paths. \textbf{This motivates the first challenge: activating reasoning requires learnable inference traces, such as Chain-of-Thought (CoT)~\citep{2022cot, 2025r2rec, 2024gpt4o}, which can structure historical evidence, extract preference cues, and compare candidates, while also providing more stable training signals for GRPO~\citep{2025-sft-2, 2025-does, 2025vision-r1}.}

To generate more accurate and informative multimodal CoT trajectories~\citep{2024RLAIF-V, 2025r1-r1}, data construction may inadvertently introduces signals related to the next item profile~\citep{2025insight-v, 2025IRLLRec}, which can create an illusion of inflated rewards during training~\citep{2026Shortcuts}. As shown in Figure~\ref{fig:intro} (b), this may inflate training rewards but hurt test-time generalization. \textbf{This raises the second challenge: the model may take shortcuts by using such cues in the reasoning text, creating a reward inflation that harms generalization.}

To tackle these issues, we propose MLLMRec-R1, a simple yet effective reasoning paradigm that integrates CoT construction with reinforcement learning (RL)~\citep{2022rlhf, 2025-sft-2} for sequential recommendation. For \textbf{Challenge 1}, we compress visual signals offline into fine-grained textual descriptions, avoiding expensive visual-token computation during training and inference while retaining multimodal semantic gains. Futhermore, we construct a high-quality multimodal CoT recommendation dataset: MLLMRec-R1 uses an existing MLLM to process image–title sequences and generate pseudo-CoT texts with explicit visual grounding and stepwise reasoning that explains interactions in a textual format. We then feed the rich reasoning texts and item descriptions into a text-only reasoning model DeepSeek-R1~\citep{2025-deepseek-r1} to extract higher-quality CoTs. For \textbf{Challenge 2}, we propose a mixed-grained data augmentation that mixes high-confidence CoT samples with standard samples to reduce shortcut learning. Specifically, we use an MLLM to embed title–image pairs and compute modality consistency scores to estimate input quality. We further measure prediction consistency between the CoT text and the predicted item to assess output reliability. Using both signals, we select high-confidence CoT samples and inject them into the training stream. \textbf{In a nutshell, MLLMRec-R1 enables MLLM-based multimodal recommendation within an GRPO pipeline: textualized visual cues improve efficiency, and a mixed-grained training strategy mitigates reward inflation.}

\begin{itemize}[leftmargin=*]
    \item We investigate the application of MLLMs to sequential recommendation and construct high-quality multimodal CoT data by compressing visual signals while preserving multimodal semantics, providing reliable supervision for reinforcement learning.
    \item We further propose a mixed-grained data augmentation that combines confidence-based samples filtering with random mixing to reduce speculative behavior in RL training.
    \item Extensive experiments on three benchmark datasets demonstrate that our approach consistently outperforms existing baselines. Furthermore, we provide a thorough analysis of the contributions of both the GRPO-based CoT construction and the mixed-grained strategy to MSR performance.
\end{itemize}

\section{Preliminary}
In this section, we first introduce the problem of general sequential recommendation, the present how we integrate LLMs to empower sequential recommendation through SFT and GRPO manners.

\subsection{Problem Formulation}
In a general sequential recommendation scenario, we consider a user set $\mathcal{U}$ and an item set $\mathcal{V}$, where $|\mathcal{U}|$ and $|\mathcal{V}|$ denote the total number of users and items, respectively. The goal is to model dynamic user interests over time by leveraging their historical interaction behaviors. Formally, given an interaction sequence $S^u = \{v^u_1, v^u_2, \cdots, v^u_t\}$ for a user $u \in \mathcal{U}$, where $v^u_t$ is the item interacted with at timestamp $t$, and $t$ indicates the sequence length, the objective is to learn a recommendation function $f_\theta$ that predicts the next item $v^u_{t+1}$ the user is most likely to engage with.

We formulate sequential recommendation with a interaction history $S^u$. To build an LLM instruction, we take the last $K$ interactions as context and construct a candidate set of size $k$, $\mathcal{C}^u=\{v^u_{t+1}\}\cup\mathcal{N}^u$, where $\mathcal{N}^u$ contains sampled negatives from non-interacted items. We then serialize $(S^u, \mathcal{C}^u)$ into an \textbf{instruction prompt} with four components: (i) a role statement, (ii) a recently watched list, (iii) a candidates list, and (iv) a strict output constraint. As illustrated in Figure~\ref{fig:intro} (a), each item is represented by a canonical token [ITEM id] and its movie title, and the \textbf{output response} is constrained to follow this format. 
Next, we present a two-stage training paradigm with SFT and GRPO. We first use SFT to adapt a pretrained LLM to follow recommendation instructions, and then apply GRPO to sample multiple responses under the same prompt and update the policy using relative advantages, further improving ranking decisions in sequential recommendation.

\subsection{Supervised Fine-Tuning (SFT)}
For a text-only LLM $\pi$, user preference learning typically begins with an instruction-tuning stage, resulting in a fine-tuned model $\pi_{\phi}$~\citep{2023tallrec}. It initializes a stable policy distribution for subsequent GRPO-based post-training. Specifically, this stage adapts a pretrained LLM into a recommendation oriented instruction follower and optimizes it on an dataset $\mathcal{D} = \left\{\left(x^u,\, y^u\right)\right\}_{u=1}^{|\mathcal{D}|}$, where $x$ denotes the \textbf{instruction prompt} and $y$ denotes the \textbf{output response}. Formally, SFT maximizes the likelihood of generating the target response $y$ conditioned on the prompt $x$:
\begin{equation}
\mathcal{L}_{\text{SFT}} = -\mathbb{E}{(x, y) \sim \mathcal{D}} \left[ \log, \pi_{\phi}(y \mid x) \right].
\label{eq1}
\end{equation}

\begin{figure*}[t]
    \centering
    \includegraphics[width=\linewidth]{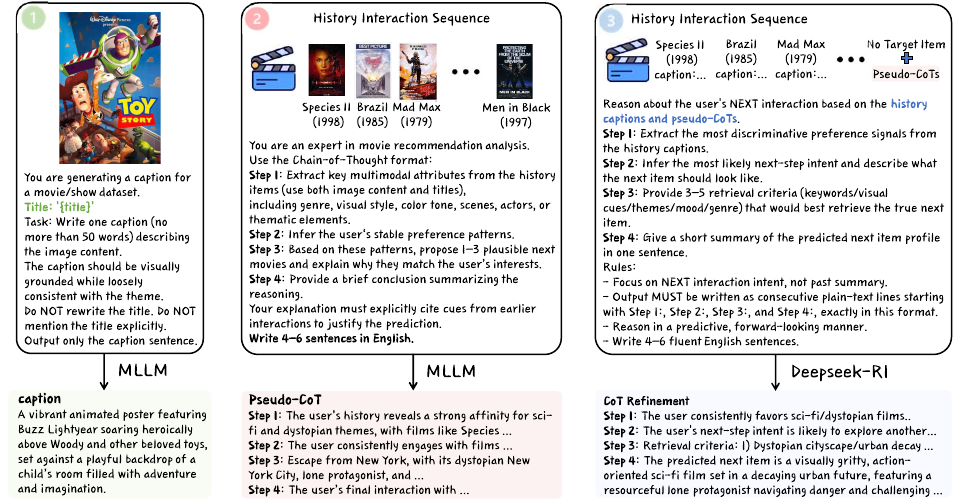}
    \caption{Multimodal CoT construction pipeline. (1) Caption generation. \textnormal{We use an MLLM to produce fine-grained captions for each item image, compressing costly visual signals into reusable text offline.} (2) Pseudo-CoT construction. \textnormal{Given the interaction history, we prompt the MLLM to generate structured CoT rationales that surface key visual cues in text beyond plain captions.} (3) CoT refinement. \textnormal{We combine these visual cues with the reasoning trace and input them to DeepSeek-R1~\citep{2025-deepseek-r1} to obtain high-quality multimodal CoT supervision.}}
    \label{fig:data}
\end{figure*}

\subsection{Group Relative Policy Optimization (GRPO)}
After SFT, we further improve the instruction-following recommender $\pi_{\phi}$ via reinforcement learning, obtaining a policy denoted as $\pi_\theta$. Different from pairwise preference optimization~\citep{2023dpo} that relies on constructing explicit chosen--rejected pairs, GRPO directly optimizes the model with \textbf{group-wise}~\citep{2024grpo} sampled responses under the same prompt. Given an \textbf{instruction prompt} $x$, we sample a group of $G$ responses $\{o_1, o_2, \ldots,o_{G}\}\sim \pi_{\theta_{\mathrm{old}}}(\cdot\mid x)$ and score each response with a task reward $r(x,o_i, y)$. GRPO uses the relative advantage within the group as the training signal:
\begin{align}
    \mathcal{J}_{\mathrm{GRPO}}(\theta)
    &=\mathbb{E}_{x \sim P(Q),\, \{o_i\}_{i=1}^{G} \sim \pi_{\theta_{\mathrm{old}}}(O|q)}\Biggl[\frac{1}{G}\sum_{i=1}^{G}\,\frac{1}{|o_i|}\sum_{t=1}^{|o_i|}
    \notag\\[-2pt]
    &\min\Biggl\{\underbrace{\frac{\pi_\theta\!\bigl(o_{i,t}\mid x,\,o_{i,<t}\bigr)}{\pi_{\theta_{\mathrm{old}}}\!\bigl(o_{i,t}\mid x,\,o_{i,<t}\bigr)}}_{\mathclap{\scriptsize\text{policy ratio}}}\hat{A}_{i,t},
    \notag\\[-2pt]
    &\underbrace{\mathrm{clip}\!\left(\frac{\pi_\theta\!\bigl(o_{i,t}\mid x,\,o_{i,<t}\bigr)}{\pi_{\theta_{\mathrm{old}}}\!\bigl(o_{i,t}\mid x,\,o_{i,<t}\bigr)},\,1-\varepsilon,\,1+\varepsilon \right)}_{\mathclap{\scriptsize\text{clipped ratio}}}\Biggr\}\hat{A}_{i,t}\Biggr]
    \notag\\[-2pt]
    &\quad -\,\beta\,\underbrace{D_{\mathrm{KL}}\!\left[\pi_\theta \,\|\, \pi_{\phi}\right]}_{\mathclap{\scriptsize\text{stability term}}}
    \label{eq2}
\end{align}
where $P(Q)$ denotes the empirical distribution of \textbf{instruction prompts}. $\pi_{\theta_{\mathrm{old}}}$ is the frozen policy used to sample ${o_i}$, and $\pi_\theta$ is the updating policy. Meanwhile, $\varepsilon$ controls the PPO-style~\citep{2017ppo} clipping range for stable updates, and $\beta$ weights the KL divergence that regularizes $\pi_\theta$ toward the SFT reference policy $\pi_{\phi}$. Moreover, we compute the token-level advantage $\hat{A}{i,t}$ from the sequence-level reward using a group mean baseline: $\hat{A}_{i,t}=r(x,o_i, y)-\frac{1}{G}\sum_{j=1}^{G}r(x,o_j, y)$ for $t\in{1,\ldots,|o_i|}$, and apply the same $\hat{A}_{i,t}$ to all tokens in $o_i$. \textbf{Finally, the design of the task reward function $r(x,o_i,y)$ will be detailed in the next section.}

\section{Methodology}
Our MLLMRec-R1 focus on high-quality training signals to stabilize GRPO post-training. In the offline stage, we use an MLLM to compress cover images into fine-grained captions and generate sequence-level pseudo-CoTs. We then perform CoT refinement with a stronger text reasoning model without providing the target item, avoiding label leakage. In the online stage, we filter low-confidence samples based on consistency and mix a small proportion of high-quality CoTs with standard data for augmentation. Finally, we adopt lightweight reward rules to support group-relative updates in GRPO, improving ranking performance and generalization.

\subsection{Multimodal Chain-of-Thought (CoT)}
Although GRPO can activate latent reasoning without large-scale CoT supervision, it is highly sensitive to the quality of training signals, especially under multimodal noise~\citep{2025vision-r1}. Therefore, we describe how to construct high-quality MSR reasoning data to support GRPO-based post-training. Specifically, our fully automated pipeline has three steps, as shown in Figure~\ref{fig:data}. First, for each item $v$, we use an MLLM (e.g., Qwen-VL~\citep{2025Qwen2.5-VL}) to convert its visual content into a fine-grained caption aligned with the cover. Let the multimodal input be $m_v = \{ I_v, T_v \}$, where $I_v$ is the cover image and $T_v$ is the title. The caption is generated by the MLLM as:
\begin{equation}
    c_v \sim p_{\theta}\!\left(c \mid m_v\right) = p_{\theta}\!\left(c \mid I_v,\, t_v\right),
    \label{eq3}
\end{equation}
where $c_v$ is the textual representation of item $v$, and $p_{\theta}$ is the MLLM's conditional generation distribution.

Next, we construct \textbf{Pseudo-CoT} to expose sequence-level multimodal preference cues \emph{without using any target-item information}. Given only the multimodal context of the interaction history $\{m_{v^u_1}, m_{v^u_2}, \ldots, m_{v^u_t}\}$, we use a prompt template with structured step constraints to guide the MLLM to generate a step-by-step reasoning trajectory:
\begin{equation}
    r^{u} \sim p_{\theta}\!\left(r \mid m_{v^{u}_{1:t}},\, \Omega_{\mathrm{step}}\right),
    \label{eq4}
\end{equation}
where $r^{u}$ denotes the pseudo-CoT text for user $u$, and $\Omega_{\mathrm{step}}$ represents the step-wise constraints imposed on the reasoning process (Part~2 of Figure~\ref{fig:data}). Pseudo-CoT is not meant to mimic complex human reasoning; instead, it serves as a modality-bridging step in data construction. It guides the MLLM to produce richer history-conditioned descriptions, reducing information loss when converting raw multimodal inputs into text representations~\citep{2025insight-v}.

Finally, \textbf{CoT refinement} inputs the captions and pseudo-CoT into a stronger text-only reasoning model to denoise, complete, and strengthen the reasoning trace, yielding high-quality supervision for GRPO. In this step, we still avoid any target-item exposure and prompt DeepSeek-R1 to generate a refined CoT for next-intent prediction using only the textualized history $\{c_{v^u_1}, c_{v^u_2}, \ldots, c_{v^u_t}\}$ and the pseudo-CoT $r_u$ (Part~3 in Figure~\ref{fig:data}). Formally, let $p_{\phi}$ denote the generation distribution of DeepSeek-R1. The refined CoT is generated as:
\begin{equation}
    \tilde{r}^{u} \sim p_{\phi}\!\left(\tilde{r} \mid c_{v^{u}_{1:t}},\, r^{u},\, \tilde{\Omega}_{\mathrm{step}}\right),
    \label{eq5}
\end{equation}
where $\tilde{\Omega}_{\mathrm{step}}$ indicates the constraints and rules of the third step. This design prevents post-hoc attribution ``explanations'' that depend on a known $v^{u}_{t+1}$, and yields finer-grained preference cues than the original pseudo-CoT.

\textbf{A natural question is why this construction works, and whether simpler alternatives could achieve comparable results.} We conduct ablation studies on the three components of CoT, and the results in Table~\ref{tab:ablation} show that removing any component leads to performance degradation. 
To further control spurious supervision signals, we enforce a target-free constraint in pseudo-CoT construction and apply an additional sanitization step during CoT refinement to filter any accidental mentions of item identifiers or future interactions (e.g., movie titles/IDs that could reveal the next item), if present. 
\textbf{It is worth emphasizing that we do not claim our CoT construction pipeline to be the only or the optimal choice. Rather, we treat high-quality multimodal CoT data construction as a crucial prerequisite for GRPO-style post-training~\citep{2025vision-r1, 2025insight-v}.} 
Specifically, in Appendix~\ref{D} we propose an automated pipeline for CoT data quality assessment. As illustrated in Figure 1, we treat each generated reasoning text as an evaluation instance and score it along five predefined dimensions: modality consistency, prediction consistency, signal density, leakage risk, and coverage hardness. We require the evaluator to output an integer rating in $[1,3, 5]$ and one evidence sentence that cites only the original reasoning text. Experimental results show that CoT refinement yields better multimodal alignment and reasoning consistency, while substantially reducing potential label leakage risk. This evaluation protocol is directly applicable to CoT data generated by other sources strategies.

\begin{figure*}[t]
    \centering
    \includegraphics[width=\linewidth]{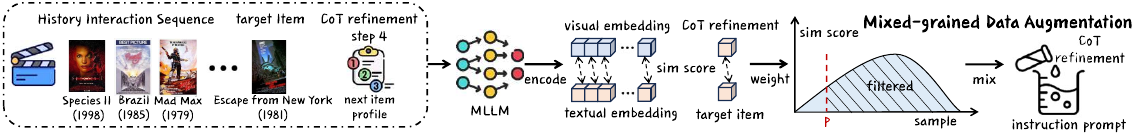}
    \caption{we compute modality consistency between title–image pairs and prediction consistency between the refined CoT profile and the target item, then mix filtered CoT refinement into the instruction data.}
    \label{fig:model}
\end{figure*}

\subsection{Mixed-grained Data Augmentation}
In this section, we first present a filtering scheme that uses \textbf{(i) modality consistency} and \textbf{(ii) prediction consistency} to remove low-quality reasoning samples, thereby reducing noisy gradients in GRPO updates. Specifically, for each interacted item $v$ in user $u$'s history sequence $S^u$, we map its multimodal input $m_v=\{I_v, T_v\}$ to a semantic embedding $\mathbf{e}_v, \mathbf{z}_v = f_m(I_v, T_v)$. We then compute the mean similarity among title–image pairs:
\begin{equation}
    \mathbf{s}_u^{'}=\frac{1}{t}\sum_{i=1}^{t}\mathrm{sim}\!\left(\mathbf{e}_{v_i^{u}},\, \mathbf{z}_{v_i^{u}}\right),
    \label{eq6}
\end{equation}
where $\mathrm{sim}(\cdot,\cdot)$ is cosine similarity. Intuitively, lower modality consistency suggests a higher risk of title ambiguity, visual noise, or cross-modal mismatch, which can degrade reasoning and learning in subsequent training. Next, we measure prediction consistency. For each sample, we denote the predicted next-item profile from Step 4 of the refined CoT $\tilde{r}^u$ as $\hat{r}^u$. We encode it as a text embedding $\mathbf{r}^u = f_m(\hat{r}^u)$ and define prediction consistency as: $\mathbf{s}_u^{''} = \mathrm{sim}(\mathbf{z}_{v_{t+1}^{u}},\mathbf{r}^u)$. Prediction consistency measures whether the reasoning trajectory matches the final prediction. If the predicted next-item profile in the CoT is inaccurate, we treat the sample as low-confidence. We simply combine the two similarity signals into a sequence-level score:
\begin{equation}
    \tau_p=\mathrm{Quantile}\!\left(\{\mathbf{s}_u\}_{u\in\mathcal{U}},\, p\right), \qquad \mathbf{s}_u = \sigma (\mathbf{s}_u^{'}+\mathbf{s}_u^{''})
    \label{eq7}
\end{equation}
where $\sigma(\cdot)$ is the sigmoid function, and $p$ denotes the retention ratio. Finally, we rank all samples by the sequence-level score $\mathbf{s}_u$  and retain those with scores higher than $\tau_p$ as high-confidence data $\mathcal{R} = \{\tilde{r}^{1}, \tilde{r}^{2}, \ldots, \tilde{r}^{k}\}$ for subsequent training. \textbf{Modality consistency to ensure reliable inputs and prediction consistency to ensure reliable outputs. Together, they filter noisy learning signals caused by cross-modal mismatch and hallucinated reasoning.} 

After filtering, we construct a mixed-grained training dataset $\mathcal{D}_{\mathrm{mix}} = \left\{\left(x^u,\, y^u,\, \tilde{r}^u\right)\right\}_{u=1}^{|\mathcal{D}|}$, where $x^u$ already includes the serialized interaction history and candidate set. For each sample, we attach a refined CoT trace $\tilde{r}^u$ only if it is selected into $R$; otherwise, we set $\tilde{r}^{u}=\varnothing$ (i.e., $\tilde{r}^{u}\notin\mathcal{R}$), and the input degenerates to a standard no-CoT prompt. $p$ typically set to 0.1 and as a low mixing ratio. This mixed-grained data augmentation allows the model learn fine-grained multimodal preference cues and reasoning patterns from a small set of high-quality CoT samples, improving ranking capability. Meanwhile, keeping most standard samples reduces reliance on shortcut cues in CoT, avoiding inflated training rewards with poorer generalization and better matching the test distribution.

\begin{figure}[t]
    \centering
    \includegraphics[width=0.75\linewidth]{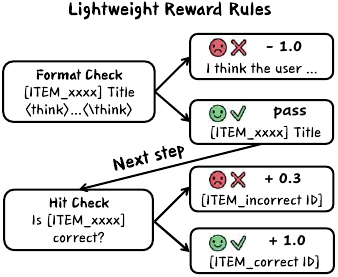}
    \caption{GRPO performs group sampling under the same context and applies reward rules with format check and hit check to provide stable training signals.}
    \label{fig:reward}
\end{figure}

\subsection{Lightweight Reward Rules}
As shown in Figure~\ref{fig:reward}, after data augmentation, we design a lightweight reward function to support GRPO’s group-relative updates. Since sequential recommendation outputs follow clear structural constraints, we decompose the reward into \textbf{format check} and \textbf{hit check}, providing stable, low-noise training signals without an extra reward model. For each response $o$ sampled under the same context, we first perform a \textbf{format check}. If $o$ does not follow the required template \texttt{[ITEM\_xxxx] Title <think> ... </think>}, we assign a penalty $R_{\mathrm{fmt}}(o)=-1.0$; otherwise, we give a base reward $R_{\mathrm{fmt}}(o)=+0.3$ to encourage valid outputs. We then perform a \textbf{hit check}: if the predicted \texttt{[ITEM\_xxxx]} matches the ground-truth next item $y$, we assign the main reward $R_{\mathrm{hit}}(o,y)=+1.0$; otherwise, the reward remains low. The final reward can be written as:
\begin{equation}
    R(o)=R_{\mathrm{fmt}}(o)+\mathbb{I}\!\left[\mathrm{hit}(o,y)\right]\cdot 1.0,
    \label{eq8}
\end{equation}
where $\mathbb{I}$ is the indicator function. The reward has $\mathcal{O}(1)$ complexity, avoiding the inference cost of an extra reward model. We allow the model to produce explanations within \texttt{<think>…</think>}, but we do not reward this part explicitly. First, free-form reasoning is hard to evaluate reliably without external verification, and rewarding it may encourage reward hacking via templated or hallucinated explanations. Second, our filtering already controls CoT quality, and GRPO mainly targets better ranking decisions. Therefore, focusing rewards on verifiable format and hit signals improves training stability and keeps optimization well aligned.

\subsection{Model Optimization with GRPO}
We train the model in two stages on the augmented data. During SFT, we apply maximum-likelihood training on the target answer $y^u$. Other SFT details follow the preliminaries and are omitted. At this step, we set the filtering ratio $p$ to 0.1, following a ``\textbf{small but high-quality}'' principle~\citep{2025-deepseek-r1, 2025vision-r1}. Before GRPO, SFT is mainly used for cold-start initialization and behavior normalization. It aligns the model with basic recommendation instructions and reduces invalid outputs that could otherwise dominate the RL stage.

We then apply GRPO. At this stage, we reduce the filtering ratio to $p=0.05$ because GRPO uses on-policy updates and is more sensitive to noisy signals. As shown in the bottom of Figure~\ref{fig:model}, for each context we sample a response set from the current policy, setting temperature and top-$p$ to $0.9$ to increase diversity and encourage exploration while keeping outputs usable. This setting yields stable convergence and consistent gains in preliminary experiments, so we do not conduct extensive tuning of temperature or top-$p$. Finally, we apply lightweight reward rules to each sampled response $o_i$ to obtain a scalar reward $r(x,o_i,y)$. Referring to the GRPO objective in the equation.~\ref{eq2}, the advantage term $\hat{A}_{i,t}$ is computed from the group-level rewards $\{r_1, r_2, \ldots, r_G\}$ corresponding to the reasoning outputs within each group. This encourages updates that responses performing better than the group average, while the KL regularization term limits deviation from the SFT reference policy. For completeness, we summarize the time complexity of each stage and provide the corresponding algorithm in Appendix~\ref{A}.

\section{Experiments}
In this section, we conduct extensive experiments and answer the following research questions:
\begin{itemize}[leftmargin=*]
\item \textbf{RQ1:} How does MLLMRec-R1 compare to the current state-of-the-art (SOTA) framework in terms of performance?
\item \textbf{RQ2:} Are the key components in our MLLMRec-R1 delivering the expected performance gains?
\item \textbf{RQ3:} What are the reasons for model's superior performance?
\item \textbf{RQ4:} How do different hyperparameters affect MLLMRec-R1?
\end{itemize}

\subsection{Experiment Setup}
\subsubsection{\textbf{Datasets}}
We evaluate our model on three multimodal datasets: Microlens~\citep{2023microlens}, Netflix~\citep{2024llmrec}, and Movielens-1M\footnote{https://grouplens.org/datasets/movielens/}. Appendix~\ref{B.1} provides details on data collection.

\subsubsection{\textbf{Baselines}}
We compare MLLMRec-R1 with representative sequential, multimodal, and LLM-based recommendation methods. The full list of baselines and their implementation details are described in Appendix~\ref{B.2}.

\subsubsection{\textbf{Metrics}}
Following the standard evaluation setting~\citep{2020lightgcn}, we adopt hit ratio (HR)~\citep{2023llamarec} and normalized discounted cumulative gain (NDCG)~\citep{2025bigrec} to evaluate model performance. Specifically, the ground-truth target item is combined with $k$ randomly sampled negative items to form $k+1$ candidates. In our experiments, $k$ is fixed to 9 by default. To further examine the model’s large-scale retrieval capability, we additionally investigate the impact of different $k$ on performance in Sec.~\ref{4.2.2}. For fair comparison, all baseline methods are evaluated using the same standardized candidate set size.

\subsubsection{\textbf{Implementations}}
To ensure fairness, both our experiments and baselines are conducted on a Linux server equipped with 8 RTX PRO 6000 GPUs. We use Qwen3-VL-8B-Instruct\footnote{https://huggingface.co/Qwen/Qwen3-VL-8B-Instruct} as the MLLM backbone for CoT data construction, and DeepSeek-R1\footnote{https://huggingface.co/deepseek-ai/DeepSeek-R1} as the large reasoning model. During the training of MLLMRec-R1, we consistently set the LoRA rank to 16, the learning rate to 1e-5, and the gradient accumulation steps to 8. During the SFT stage, the per-device batch size is set to 2: the model is trained for 3 epochs on MovieLens and Netflix, and for 5 epochs on MicroLens. During the GRPO stage, the per-device batch size is increased to 4: the model is trained for 3 epochs on MovieLens and MicroLens, and for 2 epochs on Netflix. Following the setup of~\citep{2024llara}, all datasets are split into training, validation, and test sets in an 7:1:2 ratio based on the number of sequences. Due to the limited sequence length of the datasets, all models use only the most recent 9 interactions, with the last one serving as the prediction target. Furthermore, we use the same data for training and testing all traditional baselines.

\begin{table*}[t]
\captionsetup{justification=centering}
\caption{Recommendation performance comparisons on different datasets. The superscript * indicates the Imprvement is statistically significant where the p-value is less than 0.05. Bold indicates the best results, and underline is the second-best.}
\begin{adjustbox}{width=\textwidth}
\begin{NiceTabular}[c]{c|l|c|cccc|cccc|cccc}
\toprule[1pt]
\midrule
& \multirow{2}{*}{\textbf{Model}} & \multirow{2}{*}{\textbf{Year}} & \multicolumn{4}{c|}{\textbf{Movielens-1M}} & \multicolumn{4}{c|}{\textbf{Microlens}} & \multicolumn{4}{c}{\textbf{Netflix}} \\
& & & HR@3 & HR@5 & NG@3 & NG@5 & HR@3 & HR@5 & NG@3 & NG@5 & HR@3 & HR@5 & NG@3 & NG@5 \\
\midrule
\multirow{4}{*}{\textbf{Traditional}}
&\textbf{GRU4Rec}     \citep{2016GRU4Rec}     & ICLR’16    & 0.5507 & 0.7131 & 0.4762 & 0.5236 & 0.5080 & 0.6440 & 0.4369 & 0.4756 & 0.5152 & 0.7152 & 0.3659 & 0.4494 \\
&\textbf{SASRec}      \citep{2018SASRec}      & ICDM’18    & 0.6073 & 0.7498 & 0.5090 & 0.5638 & 0.5243 & 0.6597 & 0.4439 & 0.4958 & 0.5560 & 0.7570 & 0.4042 & 0.4786  \\
&\textbf{LightGCN}    \citep{2020lightgcn}    & SIGIR’20   & 0.5808 & 0.7167 & 0.4894 & 0.5330 & 0.5138 & 0.6604 & 0.4344 & 0.4913 & 0.5430 & 0.7420 & 0.3928 & 0.4702 \\
&\textbf{CL4SRec}     \citep{2022CL4SRec}     & ICDE’22    & 0.6221 & 0.7557 & 0.5306 & 0.5861 & 0.5482 & 0.6826 & 0.4694 & 0.5186 & 0.5788 & 0.7873 & 0.4374 & 0.5091 \\
\midrule
\multirow{4}{*}{\textbf{Multimodal}}
&\textbf{LATTICE}     \citep{2021LATTICE}     & MM’21      & 0.6131 & 0.7558 & 0.5223 & 0.5762 & 0.5306 & 0.6612 & 0.4531 & 0.5036 & 0.5635 & 0.7955 & 0.4110 & 0.5038 \\
&\textbf{MoRec}       \citep{2023MoRec}       & SIGIR’23   & 0.6059 & 0.7552 & 0.5074 & 0.5642 & 0.5267 & 0.6515 & 0.4537 & 0.4970 & 0.5742 & 0.7725 & 0.4277 & 0.4887 \\
&\textbf{BM3}         \citep{2023bm3}         & WWW’23     & 0.6222 & 0.7618 & 0.5259 & 0.5788 & 0.5421 & 0.6836 & 0.4749 & 0.5291 & 0.5825 & 0.8155 & 0.4213 & 0.5358 \\
&\textbf{AB-Rec}      \citep{2025AB-Rec}      & KDD’25     & 0.6194 & 0.7513 & 0.5219 & 0.5787 & 0.5426 & 0.6740 & 0.4645 & 0.5125 & 0.5865 & 0.8094 & 0.4390 & 0.5243 \\
\midrule
\multirow{4}{*}{\textbf{LLM-based}}
&\textbf{TallRec}    \citep{2023tallrec}     & Recsys’23  & 0.6139 & 0.7324 & 0.5236 & 0.5613 & 0.5243 & 0.6597 & 0.4439 & 0.4958	& 0.5690 & 0.7885 & 0.4174 & 0.4967 \\
&\textbf{LLaRA}      \citep{2024llara}       & SIGIR‘24   & 0.6172 & 0.7490 & 0.5201 & 0.5681 & 0.5213 & 0.6799 & 0.4423 & 0.5080 & 0.5650 & 0.7700 & 0.4072 & 0.4854 \\
&\textbf{SPRec}      \citep{2025SPRec}       & WWW’25     & 0.6273 & 0.7598 & 0.5290 & 0.5738 & 0.5473 & 0.6798 & 0.4690 & 0.5138 & 0.5765 & 0.7982 & 0.4250 & 0.5167 \\
&\textbf{RecZero}    \citep{2025reczero}     & NIPS’25    & \underline{0.6595} & \underline{0.7664} & \underline{0.5639} & \underline{0.6067} & \underline{0.6032} & \underline{0.7062} & \underline{0.5166} & \underline{0.5648} & \underline{0.6520} & \underline{0.8205} & \underline{0.5195} & \underline{0.5857} \\
\midrule
\multirow{4}{*}{\textbf{MLLM-based}}
&\textbf{MSRBench}   \citep{2025MSRBench}    & WWW’25     & 0.6272 & 0.7507 & 0.5288 & 0.5743 & 0.5543 & 0.6897 & 0.4739 & 0.5258 & 0.5720 & 0.7820 & 0.4295 & 0.5029 \\
&\textbf{MLLM-MSR}   \citep{2025MLLM-MSR}    & AAAI’25    & 0.6261 & 0.7563 & 0.5313 & 0.5827 & 0.5447 & 0.6734 & 0.4655 & 0.5181 & 0.5910 & 0.8080 & 0.4429 & 0.5272 \\
\rowcolor{gray!10} &\textbf{MLLMRec-R1}         & Ours       & \textbf{0.7630*} & \textbf{0.8368*} & \textbf{0.6524*} & \textbf{0.6784*} & \textbf{0.6627*} & \textbf{0.7906*} & \textbf{0.5845*} & \textbf{0.6365*} & \textbf{0.7150*} & \textbf{0.8670*} & \textbf{0.5902*} & \textbf{0.6293*} \\
\rowcolor{gray!10} &\textbf{Improv. \%}      &            & 15.69\% & 9.19\% & 15.82\% & 11.82\% & 9.86\% & 11.95\% & 13.14\% & 12.69\% & 9.66\% & 5.67\% & 13.61\% & 7.44\% \\       
\midrule
\bottomrule[1pt]                       
\end{NiceTabular}
\end{adjustbox}
\label{tab:rec_result}
\end{table*}

\subsection{Performance Comparisons(RQ1)}
In this section, we conduct experiment under two evaluation protocols: standard candidate set and larger candidate set.
\subsubsection{\textbf{Standard Training Setting.}}
In the Metrics section, the candidate set size is 10 (one target item and nine randomly sampled negatives). Table~\ref{tab:rec_result} shows the performance improvements of MLLMRec-R1 on three public datasets. We further discuss interesting observations in the results and provide possible explanations. All experimental results
are averaged over five runs.
\begin{itemize}[leftmargin=*]
    \item {\textbf{Limitations of Traditional Methods.}}
    Sequential~\citep{2016GRU4Rec, 2018SASRec, 2022CL4SRec} and collaborative filtering~\citep{2020lightgcn} models rely on abundant interactions, while our data remains highly sparse (even with truncation on Movielens). Although CL4SRec~\citep{2022CL4SRec}, the strongest among them, uses contrastive learning-based sequence augmentation to ease the loosely defined cold-start issue, it still falls behind the LLM-based method SPRec~\citep{2025SPRec} across the three datasets, highlighting the performance bottleneck in sparse scenarios.
    \item {\textbf{Limitations of LLM-based Methods.}}
    Existing LLM-based semantic retrieval methods~\citep{2023tallrec, 2024llara, 2025SPRec, 2025reczero} typically use SFT and DPO to learn from interaction histories and sampled preference pairs. However, they often ignore the strong difficulty imbalance in sampled negatives, so DPO repeatedly reinforces contrasts dominated by easy negatives. In addition, text-only LLMs cannot leverage multimodal information like images, making it hard to address modality bias from ambiguous titles \citep{2023llmrec_fairness}. Taking RecZero~\citep{2025reczero} as an example, both RecZero and the proposed MLLMRec-R1 use GRPO for post-training. However, RecZero does not model multimodal CoT, which leads to markedly lower overall performance than MLLMRec-R1.
    \item {\textbf{Limitations of MLLM-based Methods.}}
    Existing LLM-based semantic retrieval methods~\citep{2023tallrec, 2024llara, 2025SPRec, 2025reczero} typically use SFT and DPO to learn from interaction histories and sampled preference pairs. However, they often ignore the strong difficulty imbalance in sampled negatives, so DPO repeatedly reinforces contrasts dominated by easy negatives. In addition, text-only LLMs cannot leverage multimodal information like images, making it hard to address modality bias from ambiguous titles \citep{2023llmrec_fairness}. Taking RecZero~\citep{2025reczero} as an example, both RecZero and the proposed MLLMRec-R1 use GRPO for post-training. However, RecZero does not model multimodal CoT, which leads to markedly lower overall performance than MLLMRec-R1.
    \item {\textbf{Superiority of MLLMRec-R1.}}
    Our model consistently outperforms all baselines across all metrics, with relative gains of 15.82\%, 13.14\%, and 13.61\% on MicroLens, Netflix, and MovieLens, respectively. These results indicate stronger modeling of fine-grained user preferences. The larger improvement on NG@3 further suggests better top-rank quality, with improved relevance and diversity and a partial reduction in popularity bias.
\end{itemize}

\subsubsection{\textbf{Standard Setting with More Candidates.}}\label{4.2.2}
This setting evaluates the robustness of ranking performance under larger candidate sets, reflecting more realistic and challenging retrieval scenarios. Detailed results and analysis are reported in Appendix~\ref{C.1}.

\subsubsection{\textbf{Standard Setting with Different Lengths.}}
This setting examines the effect of interaction history length on model performance, shedding light on how sequential context influences multimodal reasoning and preference learning. Detailed results are provided in Appendix~\ref{C.2}.

\begin{table*}[t]
\centering
\caption{Ablation studies of model variants on the all dataset. The superscript * indicates the Imprvement is statistically significant where the p-value is less than 0.05. Bold indicates the best results, and underline is the second-best.}
\resizebox{0.95\textwidth}{!}{
\begin{tabular}{ccccccccccccc}
\toprule
\multirow{2}{*}{\textbf{Variants}} & \multicolumn{4}{c}{\textbf{Movielens-1M}} & \multicolumn{4}{c}{\textbf{Microlens}} & \multicolumn{4}{c}{\textbf{Netflix}} \\
\cmidrule(lr){2-5} \cmidrule(lr){6-9} \cmidrule(lr){10-13}
& HR@3 & HR@5 & NG@3 & NG@5 & HR@3 & HR@5 & NG@3 & NG@5 & HR@3 & HR@5 & NG@3 & NG@5 \\
\midrule
w/o ALL               & 0.6172 & 0.7490 & 0.5201 & 0.5681 & 0.5019 & 0.6530 & 0.4283 & 0.4855 & 0.5650 & 0.7700 & 0.4072 & 0.4854 \\			
w/o GRPO              & 0.6139 & 0.7324 & 0.5236 & 0.5613 & 0.5191 & 0.6582 & 0.4397 & 0.4928 & 0.5690 & 0.8010 & 0.4174 & 0.5064 \\
w/o MDA               & 0.6465 & 0.7593 & 0.5824 & 0.6035 & 0.5647 & 0.6799 & 0.5036 & 0.5441 & 0.6505 & 0.8145 & 0.5098 & 0.5771 \\
w/o MCoT              & 0.6665 & 0.7767 & 0.5978 & 0.6287 & 0.6097 & 0.7051 & 0.5174 & 0.5625 & 0.6605 & 0.8280 & 0.5177 & 0.5899 \\
w/o CoT Refinement    & 0.6924 & 0.7791 & 0.5933 & 0.6239 & 0.6371 & 0.7558 & 0.5418 & 0.5893 & 0.6870 & 0.8249 & 0.5314 & 0.5806 \\
w/o Pseudo-CoT        & 0.7095 & 0.7864 & 0.6039 & 0.6267 & 0.6394 & 0.7526 & 0.5439 & 0.5887 & 0.6810 & 0.8280 & 0.5329 & 0.5872 \\
w/o Caption           & \underline{0.7243} & \underline{0.8021} & \underline{0.6230} & \underline{0.6435} & \underline{0.6434} & \underline{0.7623} & \underline{0.5616} & \underline{0.6109} & \underline{0.6965} & \underline{0.8415} & \underline{0.5518} & \underline{0.6067} \\
MLLMRec-R1            & \textbf{0.7630*} & \textbf{0.8368*} & \textbf{0.6524*} & \textbf{0.6784*} & \textbf{0.6627*} & \textbf{0.7906*} & \textbf{0.5845*} & \textbf{0.6365*} & \textbf{0.7150*} & \textbf{0.8670*} & \textbf{0.5902*} & \textbf{0.6293*} \\
\bottomrule
\end{tabular}}
\label{tab:ablation}
\end{table*}

\subsection{Ablation Study (RQ2)}
\subsubsection{\textbf{Impact of key components.}}
In this section, we assess the impact of key components in MLLMRec-R1 and provide potential explanations. The following model variants are evaluated:
\begin{itemize}[leftmargin=*]
\item MLLMRec-R1 w/o ALL: removes all proposed components, only SFT training paradigm is conducted;
\item MLLMRec-R1 w/o GRPO: removes GRPO-based post training, only SFT training paradigm with mixed-grained data;
\item MLLMRec-R1 w/o MDA: removes mixed-grained data agmentation, adding the all multimodal CoT into the training data;
\item MLLMRec-R1 w/o MCoT: removes multimodal CoT, only ``SFT + GRPO'' on standard instruction prompt;
\item MLLMRec-R1 w/o CoT Refinement: training using only pseudo-CoT data, without CoT refinement;
\item MLLMRec-R1 w/o Pseudo-CoT: removes pseudo-CoT when obtaining CoT refinement data;
\item MLLMRec-R1 w/o Caption: removes caption when obtaining CoT refinement data;
\end{itemize}
Table~\ref{tab:ablation} presents the experimental results on the all datasets, leading to the following insights:
Removing GRPO leads to a performance drop similar to removing all components, as both show significant degradation. This occurs because, without the RL stage for within-group preference optimization, the model cannot effectively learn relative advantages, resulting in unreliable rankings for diverse candidate sets. Although SFT can provide basic alignment and initialization, it fails to fully capture the reasoning patterns of multimodal CoT and cross-modal correlations. Next, w/o MCoT shows significantly worse performance than the full MLLMRec-R1, highlighting that text-only RL can improve recommendations but is limited without high-quality multimodal CoT data. Similarly, w/o MDA slightly underperforms w/o MCoT, indicating that, without MDA, the model tends to rely on shortcuts, using noise and superficial CoT associations, which can harm training. Overall, MDA is crucial for training stability and generalization, as it prevents biases from over-relying on CoT.

Finally, we conduct fine-grained ablations on the MCoT construction pipeline. As shown by the results, both captions and pseudo-CoT improve performance, with pseudo-CoT contributing more by providing explicit preference cues and reusable reasoning patterns, while captions mainly compress fine-grained visual semantics. Together with the phenomenon observed in Figure~\ref{fig:quality}, removing CoT refinement and training on pseudo-CoT alone sharply degrades performance, likely due to potential information conflict that hurts test-time generalization.


\begin{figure}[t]
    \centering
    \includegraphics[width=\linewidth]{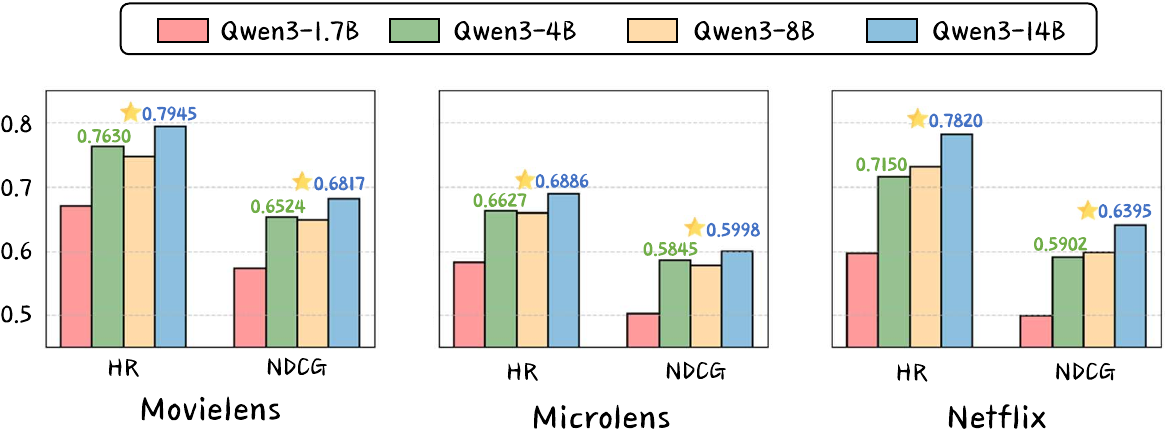}
    \caption{Performance comparison of different Backbone sizes on the all dataset  w.r.t. HR@3 and NDCG@3.}
    \label{fig:qwen}
    \vspace{-0.3cm}
\end{figure}

\subsubsection{\textbf{Impact of different LLM backbones.}}
To evaluate the impact of different MLLMs as backbones, we run experiments with various Qwen3 sizes~\citep{Qwen3-VL}. As shown in the Figure~\ref{fig:qwen}, larger-scale Qwen3 models consistently outperform smaller ones across all datasets. This suggests that larger models better leverage their reasoning capacity when enhanced with multimodal CoT data, leading to improved decision-making and ranking. In contrast, the 4B and 8B models show more mixed performance, indicating that, while they can handle multimodal reasoning to some extent, their ability to fully utilize multimodal CoT data is weaker than the larger models. This highlights the importance of model scale in maximizing the potential of multimodal Chain-of-Thought supervision, especially when combined with GRPO optimization. The weaker performance of the 1.7B model likely stems from overly strict instruction fine-tuning, which makes its outputs less aligned with the top-k candidates~\citep{2025-sft-2}. We are evaluating other LLM backbones (e.g., LLaMA~\citep{2024llama3} or Gemma~\citep{2025gemma} because they would require prompt adjustments. We expect similar trends for these models.

\subsection{Case Study (RQ3)}
To further illustrate how reasoning enhances recommendation accuracy, we provide case studies in Appendix~\ref{E}.

\subsection{Hyperparameter Sensitivity (RQ4)}
We group the hyperparameters into three sets: \textbf{GRPO-specific, reward-function, and data-filtering parameters.} \textbf{(1) GRPO-specific}. This group includes the group size $G$, sampling temperature, top-$p$ and the KL divergence coefficient $\beta$. Based on preliminary experiments, we fix temperature and top-$p$ to 0.9 to increase group sampling diversity while keeping outputs usable. \textbf{(2) Reward-function}. As discussed in the reward design section, we do not conduct extensive grid search because of this setting works well with simple validation. \textbf{(3) Data-filtering}. The filtering ratio controls the quality and noise level of samples used in GRPO and is a key factor for training stability.

\begin{figure}[t]
    \centering
    \includegraphics[width=\linewidth]{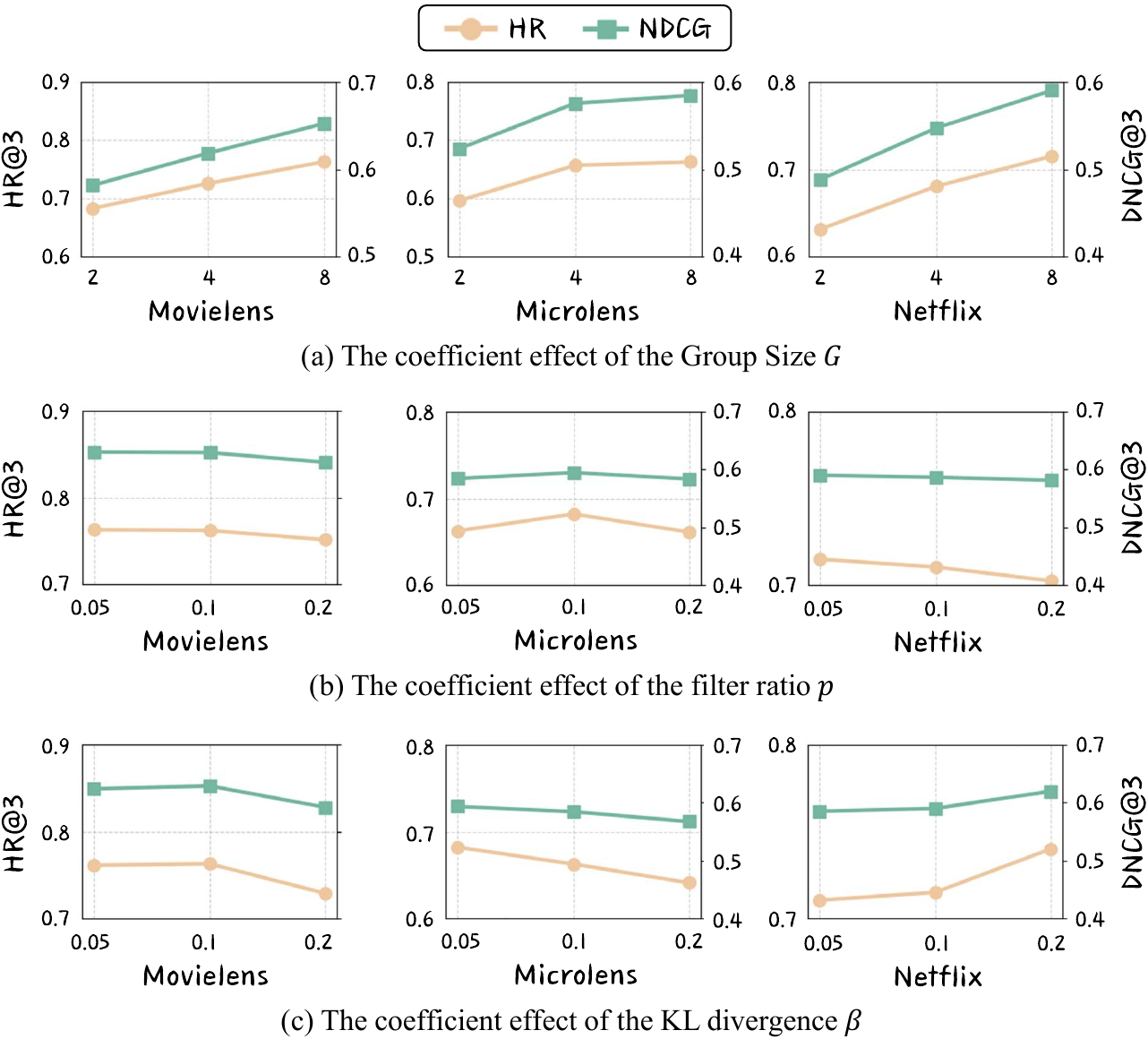}
    \caption{Performance comparison w.r.t. different hyperparameters. The $y$-axes on the left belong to the metrics HR and the right belong to NDCG.}
    \label{fig:hyperpara}
\end{figure}

Overall, our sensitivity study focuses on group size $G$, KL coefficient $\beta$, and the filtering ratio $p$ to assess their effects on training stability and final recommendation performance.
\begin{itemize}[leftmargin=*]
    \item As shown in Figure~\ref{fig:hyperpara} (a), increasing $G$ from $2$ to $8$ leads to overall gains in HR and NDCG across all three datasets, suggesting that larger groups provide more stable and informative relative-advantage signals and thus improve ranking quality.
    \item As observed from the figure, overly aggressive filtering (smaller $p$) causes slight drops on some datasets. This indicates that filtering reduces noisy gradients, but excessive removal can shrink training coverage and discard hard yet useful samples.
    \item Finally, in part (c), sensitivity to $\beta$ varies obviously across datasets, but moderate value consistently performs best. This reflects a trade-off between limiting policy drift (stability) and allowing exploratory updates (effectiveness).
\end{itemize}

\section{Related Work}
\noindent \textbf{LLMs and MLLMs for Recommendation.}
Large language models (LLMs) have gained significant attention in recommender systems for their advanced language understanding and reasoning abilities. Research in this area primarily follows three paradigms: LLMs as recommenders, enhancers, and encoders. LLMs as recommenders \citep{2024llara, 2023llamarec, 2026reinforced}: These methods use user interaction histories as prompts to guide LLMs in selecting recommendation targets from a candidate set. TallRec \citep{2023tallrec} fine-tunes Llama on constructed recommendation datasets, enhancing LLM decision-making in recommendations. RosePO \citep{2024rosepo} employs smoothing personalized preference optimization to fine-tune LLMs, improving performance while ensuring the recommender remains ``helpful and harmless''. RecZero \citep{2025reczero} trains LLMs for recommendation via reinforcement learning with structured reasoning prompts. LLMs as enhancers \citep{2025IRLLRec, 2024DALR, 2025DMRec}: RLMRec \citep{2024rlmrec} introduces a framework leveraging LLM-driven representation learning, with contrastive and generative alignment methods to improve recommendations. AlphaRec \citep{2024alpharec} replaces ID-based embeddings with language embeddings and combines GCN and CL for a simple yet effective recommendation approach. LLMs as encoders \citep{2023chatrec, 2024BLaIR}: EasyRec \citep{2024easyrec} leverages collaborative information and textual data from users and items to retrain language models for recommendation, achieving impressive performance in zero-shot scenarios. To extend LLM-based recommenders to multimodal scenarios, researchers explore multimodal large language models (MLLMs) that integrate visual and textual information to better capture user preferences. Rec-GPT4V~\citep{2024Rec-gpt4v} introduces a visual summarization reasoning mechanism to improve user modeling and reduce dynamic image redundancy. MSRBench~\citep{2025MSRBench} conducts a systematic evaluation of different MLLM integration strategies and highlights computational inefficiency as a key challenge. MLLM-MSR~\citep{2025MLLM-MSR} designs a two-stage user modeling framework that combines image summarization with sequence modeling to fine-tune MLLMs.

\noindent \textbf{LLMs Post-training.}
As a widely used post-training method, supervised fine-tuning (SFT)~\citep{2025-sft-1, 2025collm} applies maximum-likelihood learning to instruction–demonstration data, enabling stable instruction following and basic reasoning formats. However, \citep{2025-sft-2} indicates that SFT often memorizes training examples, leading to weak generalization to out-of-distribution settings, motivating reinforcement learning (RL)~\citep{2025-DAPO, 2025-does, 2025-deepseek-r1}–based post-training for better domain generalization. Proximal policy optimization (PPO)~\citep{2017ppo} improves a policy by maximizing expected reward while limiting drift from a reference model, and is effective for strengthening reasoning~\citep{2025-sft-1}. Building on this RL foundation, recent work mainly follows two lines: (1) Direct preference optimization (DPO)-based methods~\citep{2024simpo, 2025-alphadpo, 2024-kto, 2024mdpo} and (2) Group relative policy optimization (GRPO)-based methods~\citep{2024grpo, 2025-GSPO, 2025-DAPO, 2025-DrGRPO}. DPO learns from paired preferences, avoiding explicit reward modeling and in-policy rollouts as in PPO, which improves stability and efficiency. SimPO~\citep{2024simpo} further reparameterizes the objective to learn directly from preference data without training a separate reward model. AlphaDPO~\citep{2025-alphadpo} strengthens preference modeling and training robustness to reduce bias from noisy preferences and uneven sample difficulty. In contrast, GRPO~\citep{2024grpo} follows PPO’s on-policy RL setting but computes group-wise relative advantages from multiple samples under the same prompt, which better supports reasoning-oriented optimization~\citep{2025-deepseek-r1}. DAPO~\citep{2025-DAPO} further improves stability and sample efficiency by decoupling clipping and dynamic sampling. GSPO~\citep{2025-GSPO} moves importance weighting from token level to sequence level, reducing instability from token level ratio estimation.

\section{Conclusion} 
In this paper, we investigate how to effectively incentivize reasoning capability in large language models for multimodal sequential recommendation. We propose MLLMRec-R1, a unified framework that couples GRPO-based post-training with high-quality multimodal CoT supervision. It textualizes visual cues offline to reduce visual-token overhead, constructs high-quality multimodal CoT supervision, and applies mixed-grained data augmentation to filter noisy trajectories and mitigate shortcut learning. With lightweight reward rules, MLLMRec-R1 achieves consistent gains on MovieLens, MicroLens, and Netflix, demonstrating its effectiveness.

\bibliographystyle{ACM-Reference-Format}
\bibliography{sample-base}
\clearpage
\appendix
\section*{Appendix}
\addcontentsline{toc}{section}{Appendix}
\section{Time complexity analysis}\label{A}
Our framework consists of an offline multimodal CoT construction stage and an online post-training stage. The offline phase generates $N$ data points through mixed-grained  data augmentation. This invocation is a one-time offline process with minimal impact on training efficiency. In the online stage, SFT on the mixed-grained dataset follows standard instruction tuning, with complexity $\mathcal{O}(NT^2 d)$, where $T$ is the training sequence length and $d$ is the backbone hidden size. Finally, GRPO is the main computational bottleneck: for each prompt we sample $G$ outputs with average length $|o|$, and the on-policy update requires sampling plus a forward--backward pass, yielding $\mathcal{O}(N G |o|^2 d)$ in practice. Our lightweight reward rules are constant-time string/ID operations per sample, adding only $\mathcal{O}(NG)$ overhead, which is negligible compared with the Transformer term. Therefore, the overall complexity of MLLMRec-R1 can be summarized as: $\mathcal{O}\Big({N T^2 d} + {N\, G\, |o|^2 d}\Big)$.

\begin{algorithm}[h!]
\caption{Algorithm of MLLMRec-R1}
\begin{algorithmic}[1]
\item[] \textbf{Input:} user sequences $S^u$, target item $v^u_{t+1}$, item metadata $m_v=\{I_v,T_v\}$, mixed-grained set $\mathcal{D}_{\mathrm{mix}}=\{(x^u,y^u,\bar r^u)\}$, policies $\pi_\theta$ and $\pi_{\mathrm{ref}}$, group size $G$, decoding parameters (temperature, top-$p$), and reward rule $R(\cdot)$.
\item[] \textbf{Output:} CoT refinement $\widetilde{\mathcal{R}}=\{\tilde r^u\}$ and optimized policy $\pi_{\theta}$.
\STATE \textbf{Stage I: Multimodal CoT construction}
\FOR{each item $v\in\mathcal{V}$}
    \STATE Construct the captions $c_v$. \hfill $\triangleright$ {Equ~\eqref{eq3}}
\ENDFOR
\FOR{each user $u\in\mathcal{U}$ with history $S^u$ and target $v^u_{t+1}$}
    \STATE Obtain pseudo-CoT $r^u$. \hfill $\triangleright$ {Equ~\eqref{eq4}}
    \STATE Mask target information in $r^u$.
    \STATE Obtain CoT refinement $\tilde r^u$. \hfill $\triangleright$ {Equ~\eqref{eq5}}
\ENDFOR
\STATE \textbf{Stage II: GRPO post-training}
\FOR{each training sample $(x^u,y^u,\bar r^u)\in\mathcal{D}_{\mathrm{mix}}$}
    \STATE Sample $G$ responses $\{o_i\}_{i=1}^{G}\sim \pi_\theta(\cdot\mid x^u,\bar r^u)$ with (temperature, top-$p$).
    \STATE Compute rewards $\{r_i\}_{i=1}^{G}$ using $r_i=R(o_i,y^u)$. \hfill $\triangleright$ {Equ~\eqref{eq8}}
    \STATE Compute group-relative advantages $\{\hat A_i\}_{i=1}^{G}$ and GRPO loss with KL regularization to $\pi_{\mathrm{ref}}$. \hfill $\triangleright$ {Equ~\eqref{eq2}}
    \STATE Update $\pi_{\theta}$ by minimizing the GRPO loss.
\ENDFOR
\end{algorithmic}
\label{alg:mllmrec-r1}
\end{algorithm}

\section{Details of experiment setup}\label{B}
\subsection{\textbf{Datasets}}\label{B.1}
We evaluate our model on three multimodal datasets. A statistical overview of all the datasets is presented in Table \ref{table:dataset}. 
\begin{itemize}[leftmargin=*]
\item \textbf{Microlens}~\citep{2023microlens} is a short-video recommendation dataset that includes titles, cover images, and video information. In this work, only titles and images are used for MLLM. The average number of user interactions is 8.79.
\item \textbf{Netflix}~\citep{2024llmrec} comes from the data collected on Kaggle by LLMRec and is a user-based movie recommendation dataset. Nonetheless, the dataset is highly sparse, with each user interacting with only 5.23 items on average.
\item \textbf{Moivelenes}~\citep{2024DHGPF} is a classic movie recommendation dataset, where each movie includes a title, year, and corresponding poster. It is very dense, with an average sequence length of 165.56.
\end{itemize}

\subsection{\textbf{Baselines}}\label{B.2}
Our MLLMRec-R1 employs the Qwen as the backbone to serve as a sequential recommender. Therefore, we divide the baselines into traditional, multimodal, LLM-based, and MLLM-based recommendation methods.

\noindent\textbf{Traditional RS.}
GRU4Rec \citep{2016GRU4Rec} uses GRU \citep{2014GRU} to construct a session-level sequential model for next-click prediction.
SASRec \citep{2018SASRec} models the user's historical behavior sequence using a self-attention mechanism.
LightGCN \citep{2020lightgcn} removes the feature transformation and nonlinear activations of GCN for light recommendation.
CL4SRec \citep{2022CL4SRec} is the first to introduce contrastive learning into SR.

\noindent\textbf{Multimodal RS.}
LATTICE \citep{2021LATTICE} constructs a latent item–item graph to enhance multimodal recommendation performance.
MoRec \citep{2023MoRec} explores whether pure modality information can replace traditional item ID representations.
BM3 \citep{2023bm3} proposes a self-supervised multimodal model that learns representations without negative sampling.
AB-Rec \citep{2025AB-Rec} aligns ID and multimodal representations to enable effective multimodal recommendation.

\noindent\textbf{LLM-based RS.}
TallRec \citep{2023tallrec} proposes supervised fine-tuning of large language models to adapt them for recommendation tasks.
LLaRA \citep{2024llara} designs a recommendation assistant that combines collaborative filtering signals with language models.
SPRec \citep{2025SPRec} leverages a self-play mechanism in LLMs to de-bias and guide the model toward more balanced outputs.
RecZero \citep{2025reczero} trains LLMs for recommendation via reinforcement learning with structured reasoning prompts.

\noindent\textbf{MLLM-based RS.}
MSRBench \citep{2025MSRBench} evaluates the capability of vision-language multimodal large models in SR.
MLLM-MSR \citep{2025MLLM-MSR} fine-tunes multimodal LLMs to specialize them for SR tasks.

\begin{table}[t]
\captionsetup{justification=centering}
\caption{Statistics of the experimental datasets.}
\begin{adjustbox}{width=0.45\textwidth}
\begin{NiceTabular}{ccccc}
\toprule
\textbf{Dataset} & \textbf{\#Users} & \textbf{\#Items} & \textbf{\#Interactions} & \textbf{Density} \\
\midrule
Microlens  & 25,411 & 41,081 & 223,263   & $2.1 \times 10^{-4}$ \\
Netflix    & 13,187 & 17,366 & 68,933    & $3.0 \times 10^{-4}$ \\
Movielens  & 6,040  & 3,952  & 1,000,209 & $4.2 \times 10^{-2}$ \\
\bottomrule
\end{NiceTabular}
\end{adjustbox}
\label{table:dataset}
\vspace{-0.4cm}
\end{table}

\section{Extension of the main experiment}\label{C}
\subsection{\textbf{Standard setting with more candidates.}}\label{C.1}
To evaluate large-scale candidate retrieval, we increase the candidate set to 100 by sampling 99 negatives. \textbf{Without retraining, we directly test the model trained under the standard setting}, and report the results in Table~\ref{tab:neg}. We include several classic methods as baselines and exclude existing MLLM-based approaches, since their visual-token overhead makes inference hard to scale to large candidate sets. As shown, compared with the runner-up RecZero, MLLMRec-R1 achieves larger relative gains under larger candidate sets, especially on NDCG. This benefit stems from multimodal CoT reasoning supervision and GRPO’s group-relative optimization: the former provides fine-grained preference signals andreduces noisy updates from hallucinated or low-confidence trajectories, while the latter learns more discriminative ranking behavior and better keeps the target item near the top even with many more negatives~\citep{2025-deepseek-r1, 2025r1-r1}.

\subsection{\textbf{Standard setting with different lengths.}}\label{C.2}
We further examine how the history length $t$ affects performance on MovieLens, with results reported in Table~\ref{tab:length}. For LLM-based models, performance increases with $t$, peaks at $t=10$, and then drops, suggesting that the benefits of GRPO and CoT are not monotonic: Longer histories may add weakly related interactions and noise, yielding diminishing returns. In contrast, sequential models perform similarly at $t=10$ and $t=20$, likely because they rely more on local patterns and gain less from additional context. Moreover, setting $t=20$ removes users with shorter histories, reducing training data and weakening supervision.

\begin{table}[t]
\centering
\caption{Performance comparison with more candidates.}
\resizebox{0.45\textwidth}{!}{
\begin{tabular}{cccccc}
\toprule
\textbf{Dataset} & \textbf{Model} & HR@5 & HR@10 & NG@5 & NG@10 \\
\midrule
\multirow{5}{*}{\textbf{Movielens}}
& CL4SRec~\citep{2022CL4SRec}     & 0.1427 & 0.1991 & 0.1084 & 0.1198 \\
& TallRec~\citep{2023tallrec}     & 0.1317 & 0.1930 & 0.0930 & 0.1108 \\
& RecZero~\citep{2025reczero}     & \underline{0.1963} & \underline{0.2418} & \underline{0.1432} & \underline{0.1530} \\
& MLLMRec-R1                      & \textbf{0.2444} & \textbf{0.2983} & \textbf{0.1806} & \textbf{0.1969} \\
\cmidrule{2-6}
\rowcolor{blue!5} & Improv. \%    & 24.50\%  & 23.37\%  & 26.12\%  & 28.69\% \\
\midrule
\multirow{5}{*}{\textbf{Netflix}}
& CL4SRec~\citep{2022CL4SRec}     & 0.1080 & 0.1872 & 0.0671 & 0.0937 \\
& TallRec~\citep{2023tallrec}     & 0.0980 & 0.1790 & 0.0591 & 0.0836 \\
& RecZero~\citep{2025reczero}     & \underline{0.1963} & \underline{0.2418} & \underline{0.1432} & \underline{0.1530} \\
& MLLMRec-R1                      & \textbf{0.1715} & \textbf{0.2560} & \textbf{0.1199} & \textbf{0.1512} \\
\cmidrule{2-6}
\rowcolor{blue!5} & Improv. \%    & 38.87\%  & 15.06\%  & 56.91\%  & 41.84\% \\
\bottomrule
\end{tabular}}
\label{tab:neg}
\end{table}

\begin{table}[t]
\centering
\caption{Performance comparison with different history lengths $t$ on Movielens dataset.}
\resizebox{0.45\textwidth}{!}{
\begin{tabular}{cccccc}
\toprule
\textbf{Length} & \textbf{Model} & HR@5 & HR@10 & NG@5 & NG@10 \\
\midrule
\multirow{5}{*}{\textbf{5}}
& CL4SRec~\citep{2022CL4SRec}     & 0.5919 & 0.7492 & 0.4822 & 0.5452 \\
& TallRec~\citep{2023tallrec}     & 0.5993 & 0.7566 & 0.4935 & 0.5592 \\
& RecZero~\citep{2025reczero}     & \underline{0.6161} & \underline{0.7275} & \underline{0.5436} & \underline{0.5639} \\
& MLLMRec-R1                      & \textbf{0.7119} & \textbf{0.8353} & \textbf{0.6077} & \textbf{0.6574} \\
\cmidrule{2-6}
\rowcolor{blue!5} & Improv. \%    & 15.55\%  & 10.40\%  & 11.79\%  & 16.58\% \\
\midrule
\multirow{5}{*}{\textbf{10}}
& CL4SRec~\citep{2022CL4SRec}     & 0.6221 & 0.7557 & 0.5306 & 0.5861 \\
& TallRec~\citep{2023tallrec}     & 0.6139 & 0.7324 & 0.5236 & 0.5613 \\
& RecZero~\citep{2025reczero}     & \underline{0.6595} & \underline{0.7664} & \underline{0.5639} & \underline{0.6067}\\
& MLLMRec-R1                      & \textbf{0.7630} & \textbf{0.8368} & \textbf{0.6524} & \textbf{0.6784} \\
\cmidrule{2-6}
\rowcolor{blue!5} & Improv. \%    & 15.69\% & 9.19\% & 15.82\% & 11.82\% \\
\midrule
\multirow{5}{*}{\textbf{20}}
& CL4SRec~\citep{2022CL4SRec}     & 0.6224 & 0.7521 & 0.5469 & 0.5817 \\
& TallRec~\citep{2023tallrec}     & 0.5637 & 0.7199 & 0.4765 & 0.5391 \\
& RecZero~\citep{2025reczero}     & \underline{0.6325} & \underline{0.7455} & \underline{0.5613} & \underline{0.5958} \\
& MLLMRec-R1                      & \textbf{0.7433} & \textbf{0.8369} & \textbf{0.6404} & \textbf{0.6760} \\
\cmidrule{2-6}
\rowcolor{blue!5} & Improv. \%    & 17.52\%  & 12.26\%  & 14.09\%  & 13.46\% \\
\bottomrule
\end{tabular}}
\label{tab:length}
\vspace{-0.3cm}
\end{table}

\section{Details of data-quality evaluation}\label{D}
To assess the quality and potential risks of the constructed CoT data, we design a hybrid protocol that combines automated and human evaluation, as shown in Figure~\ref{fig:quality_prompt}. We randomly sample 200 reasoning texts from the generated CoT refinement and pseudo-CoT sets, and ask GPT-5.2~\citep{2025gpt5}, Claude-4.5~\footnote{https://claude.ai/}, and human annotators~\citep{2025GRefer} to independently rate each sample along five dimensions—modality consistency, prediction consistency, signal density, leakage risk, and coverage hardness. We aggregate the results into the radar chart in Figure 7. The key motivation is that CoT construction may introduce target-related shortcut signals, artificially inflating training rewards and harming generalization; thus, potential leakage and noise must be explicitly assessed and controlled. In addition, our filtering and confidence modeling emphasize modality and prediction consistency to mitigate noisy learning signals induced by cross-modal mismatch and hallucinatory reasoning. We find that MCoT scores higher than pseudo-CoT on all dimensions. Together with the performance gains, this suggests that higher-quality CoT supervision better strengthens the model’s reasoning ability~\citep{2025-does, 2025vision-r1}.

\begin{figure*}[t]
    \centering
    \includegraphics[width=\linewidth]{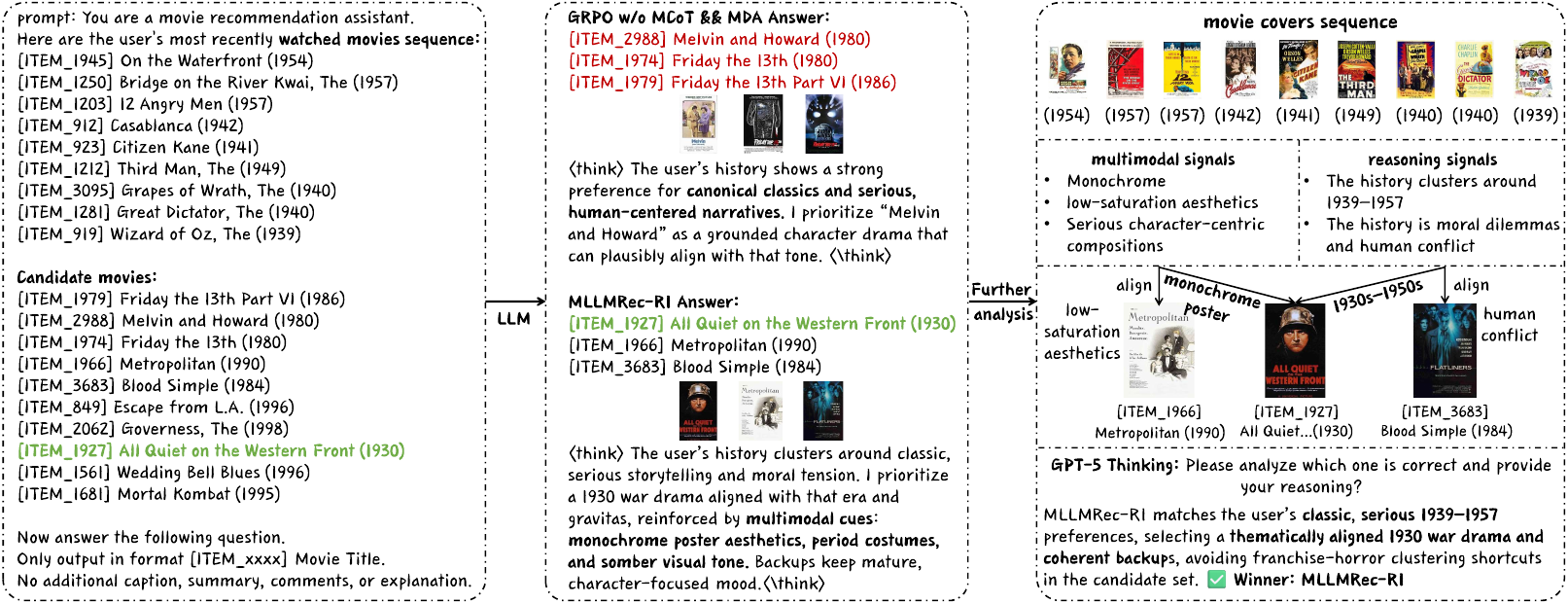}
    \caption{Case study of sequence 1127 in the MovieLens test set. ``GRPO w/o MCoT \&\& MDA'' refers to a training variant that removes multimodal CoT reasoning and the data augmentation module. The green movie is target item.}
    \label{fig:case}
\end{figure*}

\section{Case study}\label{E}
This section presents a case study of user 1127 from the Movielens dataset. Figure ~\ref{fig:case} (left) shows the inference prompt template, which includes only the interaction history and the candidate set, without explicitly adding multimodal CoT. The middle of figure compares the outputs of text-only GRPO and MLLMRec-R1 using Qwen3’s ``thinking'' format. The Top-3 results of text-only GRPO contain two movies from the same Friday the 13th horror franchise, suggesting that without multimodal CoT and MDA the model may rely on shallow co-occurrence signals (e.g., series repetition) and miss the true target item. In contrast, MLLMRec-R1 returns more diverse and semantically consistent candidates and ranks the target at Top-1. This indicates that multimodal CoT provides fine-grained cues that help MLLMRec-R1 make more reliable within-group distinctions, improving ranked recommendation quality.

Finally, the right panel offers a finer alignment analysis. Using the user’s historical cover sequence as multimodal evidence, we observe shared visual cues, including mostly monochrome and low-saturation styles and serious, character-centered compositions. These cues suggest a preference for classic period tones and morally driven human conflicts. Based on this profile, MLLMRec-R1 matches candidates to the inferred user representation and ranks All Quiet on the Western Front at Top-1. Metropolitan (1990) and Blood Simple (1984) are also ranked highly, as they fit the user’s preference for serious, character-driven stories through dialogue-focused relationships and a cold noir narrative, respectively. Overall, multimodal CoT provides fine-grained signals and, with GRPO’s within-group learning, reduces shortcuts such as franchise repetition and genre keywords, leading to more stable and generalizable rankings.

\section{Reward convergence}\label{F}
Within the first 0.2–0.5 epoch, the smoothed reward increases rapidly, showing that the policy quickly learns to align with user preferences. Afterward, gains become gradual, and the reward stabilizes around 2–3 epochs. Netflix begins with a lower reward but improves more and fluctuates more, suggesting a harder optimization problem while still benefiting strongly from post-training. In contrast, MovieLens and MicroLens converge more smoothly with smaller late-stage gains, indicating diminishing returns as the policy nears a local optimum.

\begin{figure*}[t]
    \centering
    \includegraphics[width=\linewidth]{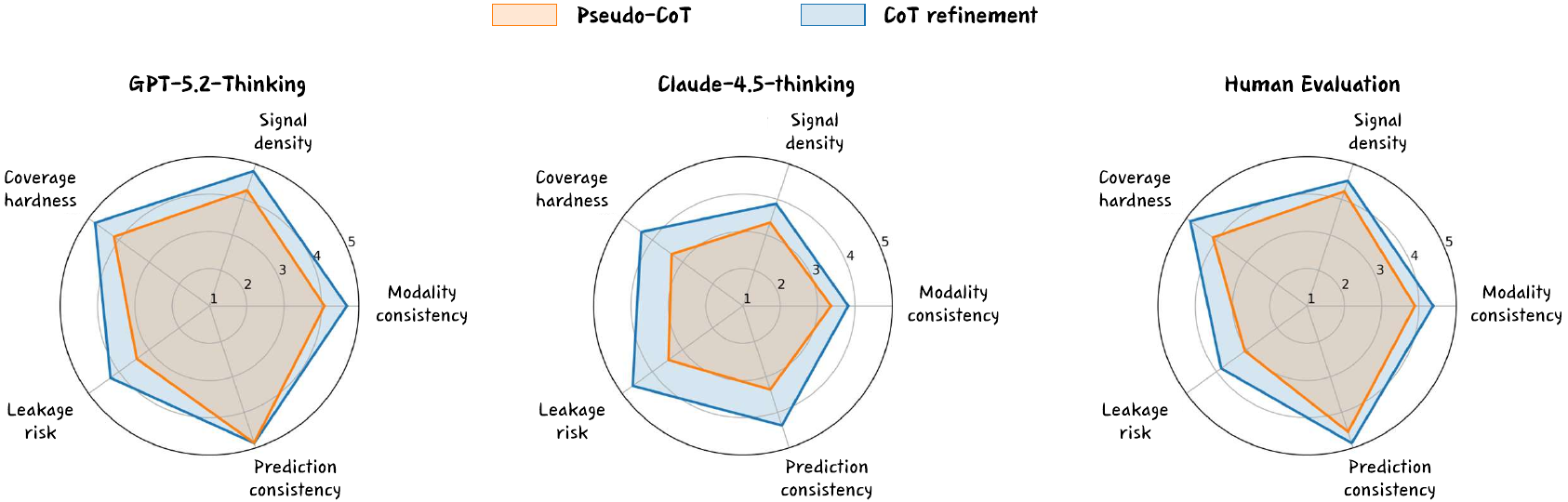}
    \caption{Radar chart comparison of Pseudo-CoT and CoT refinement on five data-quality dimensions, where scores are averaged over three datasets. We report results under GPT-5.2-thinking, Claude-4.5-thinking, and human evaluation.}
    \label{fig:quality}
\end{figure*}

\begin{figure*}[t]
    \centering
    \includegraphics[width=\linewidth]{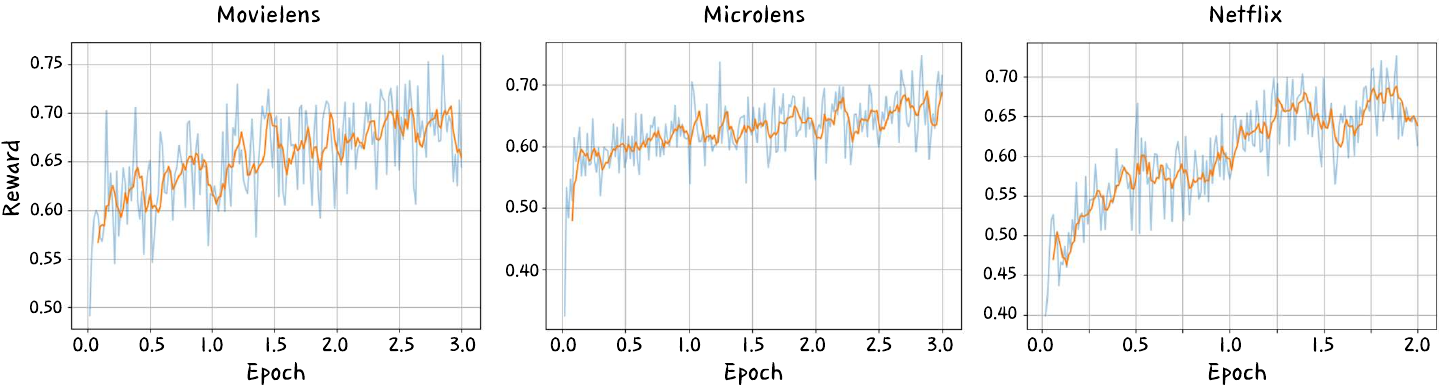}
    \caption{GRPO-based training reward curves on three benchmark datasets.}
    \label{fig:reward_value}
\end{figure*}

\begin{figure*}[t]
\centering
\begin{tcolorbox}[
  title={Prompt for Multimodal Sequential Recommendation Data-Quality Evaluation},
  colback=gray!3,
  colframe=gray!40,
  boxrule=0.5pt,
  arc=2pt,
  left=6pt,right=6pt,top=6pt,bottom=6pt
]
\small
\noindent \textbf{Role.}
You are an evaluator for ``multimodal sequential recommendation data quality.''
You will be given a Chain-of-Thought (CoT) reasoning text.
Your task is to score it on five data-quality dimensions and provide evidence for each score.

\vspace{4pt}
\noindent \textbf{Object to Evaluate.}
\begin{itemize}[leftmargin=1.2em, itemsep=1pt, topsep=2pt]
  \item \textbf{Input:} a reasoning text.
\end{itemize}

\vspace{2pt}
\noindent \textbf{Scoring Requirements.}
\begin{itemize}[leftmargin=1.2em, itemsep=1pt, topsep=2pt]
  \item For each dimension, assign an integer score from 1 to 5.
  \item For each dimension, provide exactly \textbf{ONE} sentence of evidence/justification, quoting only from the given reasoning text.
  \item Output MUST be strict JSON (no extra explanations, no markdown, no multiple blocks).
\end{itemize}

\vspace{2pt}
\noindent \textbf{Five Dimensions + Rubrics.}

\vspace{2pt}
\noindent \textbf{(1) \texttt{modality\_consistency}.}
Evaluate whether the preference signals in the reasoning are concrete and groundable to multimodal inputs (title/image/caption), avoiding vague or unverifiable claims.
\begin{itemize}[leftmargin=1.2em, itemsep=1pt, topsep=2pt]
  \item \textbf{5:} Specific and verifiable descriptions that can align with image.
  \item \textbf{3:} Generally relevant but somewhat generic; limited details; partially hard to verify.
  \item \textbf{1:} Mostly vague/template-like statements, or obviously ungroundable/contradictory descriptions.
\end{itemize}

\vspace{2pt}
\noindent \textbf{(2) \texttt{prediction\_consistency}.}
Evaluate whether steps logically support the conclusion, avoiding ``jump-to-conclusion.''
\begin{itemize}[leftmargin=1.2em, itemsep=1pt, topsep=2pt]
  \item \textbf{5:} Conclusion clearly inherits multiple retrieval criteria from steps.
  \item \textbf{3:} The main line is consistent but with missing links or weak support.
  \item \textbf{1:} Contradictions, or conclusion is unrelated to the earlier preference profile.
\end{itemize}

\vspace{2pt}
\noindent \textbf{(3) \texttt{signal\_density}.}
Evaluate whether the reasoning provides sufficiently rich, fine-grained, learnable preference cues rather than generic templates.
\begin{itemize}[leftmargin=1.2em, itemsep=1pt, topsep=2pt]
  \item \textbf{5:} Multiple fine-grained preference dimensions with strong discriminative power for candidates.
  \item \textbf{3:} Some preference cues but coarse (e.g., only ``likes sci-fi action''); moderate discriminativeness.
  \item \textbf{1:} Almost no learnable preference information, or highly templated/empty content.
\end{itemize}

\vspace{2pt}
\noindent \textbf{(4) \texttt{leakage\_risk}.}
Evaluate whether the text contains shortcut signals that enable guessing without true preference reasoning.
\begin{itemize}[leftmargin=1.2em, itemsep=1pt, topsep=2pt]
  \item \textbf{5:} No explicit target info; stays at ``profile/criteria'' level; does not uniquely pinpoint an answer.
  \item \textbf{3:} No explicit title/ID, but includes highly identifying combinations or overly strong hints.
  \item \textbf{1:} Contains target title/ID, candidate position hints, or any explicit leakage of the answer.
\end{itemize}

\vspace{2pt}
\noindent \textbf{(5) \texttt{coverage\_hardness}.}
Evaluate whether this sample provides useful difficulty and coverage for training (distinguishing similar candidates, capturing long-tail preferences), rather than being too easy or redundant.
\begin{itemize}[leftmargin=1.2em, itemsep=1pt, topsep=2pt]
  \item \textbf{5:} Fine-grained constraints that help distinguish homogeneous candidates; ``hard-but-useful'' training value.
  \item \textbf{3:} Medium difficulty and medium information value.
  \item \textbf{1:} Too easy (candidates are trivially separable) or highly repetitive, yielding low marginal training benefit.
\end{itemize}

\vspace{2pt}
\noindent \textbf{Output Format (Strict JSON).}
\begin{verbatim}
{
  "modality_consistency": {"score": 1-5, "evidence": "..."},
  "prediction_consistency": {"score": 1-5, "evidence": "..."},
  "signal_density": {"score": 1-5, "evidence": "..."},
  "leakage_risk": {"score": 1-5, "evidence": "..."},
  "coverage_hardness": {"score": 1-5, "evidence": "..."}
}
\end{verbatim}

\noindent \textbf{Now Start Evaluating.}
Please score ONLY based on the following reasoning text: \texttt{<REASONING\_TEXT>}
\end{tcolorbox}
\caption{Prompt used for automatic evaluation of multimodal CoT data quality.}
\label{fig:quality_prompt}
\end{figure*}

\end{document}